# AI and Holistic Review: Informing Human Reading in College Admissions

AJ Alvero
Stanford University
ajalvero@stanford.edu

Noah Arthurs
Stanford University
narthurs@stanford.edu

anthony lising antonio
Stanford University
aantonio@stanford.edu

Benjamin W. Domingue
Stanford University
bdomingu@stanford.edu

Ben Gebre-Medhin
Stanford University
beniam@stanford.edu

Sonia Giebel
Stanford University
sgiebel@stanford.edu

Mitchell L. Stevens
Stanford University
stevens4@stanford.edu

## ABSTRACT

College admissions in the United States is carried out by a human-centered method of evaluation known as holistic review, which typically involves reading original narrative essays submitted by each applicant. The legitimacy and fairness of holistic review, which gives human readers significant discretion over determining each applicant's fitness for admission, has been repeatedly challenged in courtrooms and the public sphere. Using a unique corpus of 283,676 application essays submitted to a large, selective, state university system between 2015 and 2016, we assess the extent to which applicant demographic characteristics can be inferred from application essays. We find a relatively interpretable classifier (logistic regression) was able to predict gender and household income with high levels of accuracy. Findings suggest that data auditing might be useful in informing holistic review, and perhaps other evaluative systems, by checking potential bias in human or computational readings.

## CCS CONCEPTS

• **Computing methodologies** → **Classification and regression trees**; • **Applied computing** → **Education**; *Sociology*.

## KEYWORDS

college admissions, fairness, bias, holistic review, supervised learning, text analysis, natural language processing, data auditing

## 1 INTRODUCTION

US selective admissions is, quite literally, on trial. In November 2014, plaintiffs filed suit against Harvard University, claiming its evaluation protocol is biased and unfair [25]. Arguments in federal court have centered on whether the evaluation protocol, known generically as holistic review, unlawfully accommodates bias in the evaluation of the non-numeric components of college applications (Students for Fair Admissions, Inc. v. President and Fellows of Harvard College, Civil Action No. 1:14-cv-14176-ADB, p. 127). Though the case has attracted attention for its potential to dismantle racial affirmative action, the lawsuit also takes aim at holistic review itself. Despite this recent challenge to its legitimacy, as well as its troubling origins in anti-Semitism [27], holistic review is recognized by admissions professionals and Supreme Court rulings as the best available process for distributing scarce academic opportunity [17].

Holistic review is premised on the notion that quantitative measures of academic accomplishment, such as GPA and standardized test scores, are insufficient bases of merit on their own and necessitate a qualitative complement [41]. As an evaluative protocol, holistic review aims to integrate multiple indicators of merit when deriving a final admission decision for each applicant. Indicators typically include scores on standardized tests, school grades, documented extracurricular accomplishments, letters of recommendation, and original essays submitted by each applicant. The purpose of these essays is not only to showcase applicants' expository skills, but also to enable evaluators' assessment of applicants as individuals and whole persons. The entire process of holistic review is predicated on the notion that human readers are essential to the task of evaluation, especially regarding the qualitative elements of applications [2].

In addition to legal challenges, the staggering number of applications to selective schools raises the cost of reliance on human readers and aggravates the potential for human bias. Stanford University, for example, must winnow over 47,000 applications down to just 2,000 undergraduate admits each year[1], a task requiring a large cadre of full-time administrators and part-time staff. Four campuses of the University of California each evaluate over 90,000 applications for admission [35]. Thus it is reasonable to suspect that computational means of "reading" large corpora of texts to extract formal regularities and outlying cases [36] may be seriously considered by admissions personnel. Already, scoring of the written portion of the GRE includes an automated computer score alongside a human one [1]. In light of escalating application numbers, cost-containment incentives, and ambitious business firms in the education technology sector, review of admissions essays is a likely domain in which computational advocates and entrepreneurs challenge the authority and cost of human-centered holistic review as the default assessment method.

---
[1] https://admission.stanford.edu/apply/selection/statistics.html

The rapid technological advance of computational reading generally raises new questions and possibilities for the evolution of holistic review. On the one hand, because computational technologies are applied to the world by human agents, they can rapidly scale implicit and explicit biases of those agents [20]. On the other hand, to the extent that computational readings can reveal demographic patterns in textual corpora [29, 39], they might sensitize human readers to risks of biased reading and inform more equitable consideration of written application materials. It is in pursuit of this latter future that we offer our analysis. Specifically, we offer preliminary investigation of how computational reading might surface demographic patterns on a key qualitative component of college applications: personal essays. We examine a novel corpus of 283,676 admission essays submitted by 93,136 self-identified Latinx applicants to a selective public research university system in 2015 and 2016. Empirically, our data audit observed the extent to which applicant gender and household income can be inferred exclusively on the basis of written applications. Using basic machine-learning models, we are able to predict gender and household income quite accurately (highest $f1$ score on gender: 79%; highest $f1$ score on income: 74%).

On the basis of this empirical finding, we make two normative arguments: one about the forward evolution of holistic review, and one about the potentially salutary role of computational readings to inform qualitative evaluation generally. First, we argue that any deployments of computational reading in holistic review should begin with the presumption that applicants' qualitative submissions carry strong signals of author demography. This means that evaluators should be cautious about using computational readings exclusively to assess applicant merit. Second, we argue that the capacity of computational reading to reveal demographic patterns in large corpora of application texts might usefully serve the improvement of qualitative assessments generally by sensitizing humans to mitigate risk of biased readings.

## 2 BACKGROUND

In this section we briefly review the history of holistic review in US selective admissions, summarize the most recent judicial scrutiny of the practice, and consider the prospect of using tools of computational reading to inform and improve the practice.

### 2.1 Historical Legacy

Holistic review arose in the United States in the early decades of the twentieth century, when a handful of elite private schools came to require a range of application materials: high school grades, scores on standardized tests, demonstrated extracurricular accomplishment, and personal essays [44]. Historians explain that the move was motivated by a demographic shift in elite universities' applicant pools: quantitative measures of fitness, specifically standardized test scores, had come to favor the admission of Jewish applicants. Status-conscious admissions officers responded by adding other elements of assessment that could be used to favor applicants from WASP families, the schools' traditional clients [27, 43]. In short, qualitative components of applications were introduced as a mechanism to legitimate the exclusion of Jewish applicants.

Yet over time – and however ironically, given the motivation of its origins – holistic review has come to be regarded as a safeguard against the bias of any single measure of fitness for college. In part, this has been due to increasingly vocal claims of bias embedded in quantitative measures of the application: that the SAT and ACT are better measures of socioeconomic status than intellectual acuity [13], and that academic potential cannot be captured fully by measures of prior academic accomplishment [38]. The hope has been that the inclusion of qualitative components in applications can provide a more comprehensive representation of each applicant's potential than quantitative measures could on their own.

### 2.2 Judicial Scrutiny

Holistic review has long been implicated in debates about the legality of giving preferential advantage to members of particular ethnoracial groups in selective admissions. In two cases involving the University of Michigan in 2003, the US Supreme Court ruled that while the use of numerical scores to advantage some applicants on the basis of race was unconstitutional (Gratz v. Bollinger, 539 U.S. 244) [22], positive consideration of race was constitutional in the context of a "highly individualized, holistic review of each applicant's file, giving serious consideration to all the ways an applicant might contribute to a diverse educational environment" (Grutter v. Bollinger, 539 U.S. 306) [24]. A subsequent Supreme Court ruling (Fisher v. University of Texas, 579 U.S.) re-affirmed the constitutionality of race-sensitive admissions evaluation under conditions of holistic review [19].

The consideration of race in admissions and the practice of holistic review remain under public and legal scrutiny. In 2014, Students for Fair Admissions, purportedly acting on behalf of Asian-American applicants, filed a civil suit against Harvard University in federal court, arguing that Harvard's admissions protocol discriminated against Asian-American applicants by systematically undervaluing their non-academic "personal" characteristics. To support their claims of bias, plaintiffs cited statistical evidence that Asian-Americans are admitted to academically selective schools at significantly lower rates than applicants with similar grades and test scores [16].

In October 2019 Judge Allison D. Burroughs ruled in favor of Harvard, declaring that "Harvard's admissions process survives strict scrutiny" (Students for Fair Admissions, Inc. v. President and Fellows of Harvard College, Civil Action No. 1:14-cv-14176-ADB, p. 127) [42]. Nevertheless, Burroughs explicitly encouraged Harvard to improve its protocols: "The process would likely benefit from conducting implicit bias trainings for admissions officers," she wrote, "...monitoring and making admissions officers aware of any significant race-related statistical disparities in the rating process" (p. 127). Noting that "now that Harvard and other schools can see how statistical analyses can reveal perhaps otherwise imperceptible statistical anomalies...statistics should be used as a check on the process and as a way to recognize when implicit bias might be affecting outcomes" (128). While Judge Burroughs' opinion upholds the base integrity of holistic review, her recommendation does question the previously unchallenged expertise of admissions personnel in judging applicants' merit. Her call for bias training may signal widespread changes in the field, as other colleges and universities shore up their admissions practices against claims of bias. Indeed, Judge Burroughs' opinion is particularly relevant as a harbinger of future admissions practice, as the Supreme Court has previously pointed to the "Harvard way" as the model holistic admissions protocol that universities can legally implement. Her implied mandate to address implicit bias through statistical analyses – in the context of escalating application numbers – portends the the use of AI and computational reading in the near future [27].

## 2.3 Computational Reading

Some of the most notable advances in AI over the past decade have been in the computational reading of texts. Unsupervised models have been shown to encode word analogies in vector space [34] and generate document topics for classification or qualitative inference [4]. Researchers have also developed methods for combining unsupervised approaches with supervised deep learning techniques to classify documents with high accuracy and precision [12]. Such methods are gradually becoming more widely used in educational research [18]. To the extent that the qualitative features of applications included in holistic review are written texts, these materials are highly amenable to computational reading.

Of course, the deployment of these techniques to an evaluation system as fateful as selective admissions is risky. Fairness and transparency concerns about AI have appropriately been raised in the context of facial recognition [7], recidivism prediction [14], and online advertising [8]. AI is often described as having the ability to rapidly scale discrimination and exacerbate social inequality; that could be the case if AI systems were to be used to adjudicate or recommend admissions evaluations or decisions. The presence of evaluative biases without modern AI is also relevant, such as the case of a computational protocol for medical school admissions that learned (from biased data) to prefer white males over other applicants [33]. This concern was expressed explicitly by Judge Burroughs in the Harvard case, when she noted that "statistics should be used as a check on the process" of holistic review, not a replacement for it. As Burroughs wrote in her opinion, "[T]he court will not dismantle a very fine admissions program that passes constitutional muster, solely because it could do better (p. 127)."

In what follows, we investigate the extent to which computational reading of applications might be used to fulfill Judge Burroughs' encouragement to use statistical analysis to improve holistic review. The ability of computational reading to reveal "otherwise imperceptible statistical anomalies" in large corpora could be a powerful tool in selective college admissions if it provides insights for fair evaluation that human readings on their own cannot discern.

## 3 DATA

Because colleges and universities carefully guard the privacy of their applicants and the confidentiality of their admissions processes, researchers rarely have the opportunity to openly perform the kind of statistical analyses called for by Judge Burroughs in the Harvard case. Our research team has been given the rare opportunity of partial access to a large corpus of applications submitted to a public university system with selective admissions. We use this opportunity to investigate the potential for computational reading to discern demographic correlates of important qualitative components of applications: personal essays.

## 3.1 The Corpus

The data for our empirical inquiry comprise 283,676 essays submitted by 93,136 applicants to a multi-campus public research university with selective admissions in the United States. The essays were submitted during the admission cycles of the 2015-2016 and 2016-2017 academic years (referred to as 2015 and 2016 respectively from here on out). These essays were required components of applications. Significantly, we also have basic demographic information describing applicants, including their self-reported household income and self-reported gender. As discussed above, income has been identified as a problematic source of bias in the evaluation of merit for college admissions. And, in natural language processing, gender is a widely understood example of how demography is embedded in text [30]. We leverage these data to assess the extent to which demographic characteristics might be inferred from computational readings of essays alone. There were other potential classification labels in the data, but we justify and explain our decision making process below.

The data were obtained as part of a larger study of Latinx young people; the entire sample self-identified as Latinx. While this prevents us from observing potential racial patterning in the corpus, it has the advantage of enabling us to eliminate between-race variation in language use as a determinant of our findings. Controlling for race in this way also reveals important patterns of intra-racial linguistic variation that otherwise would be difficult to see.

For the 2015 application cycle, students were required to write two essays in response to the same two prompts. In 2016 the application protocol was changed such that students were required to select four prompts from eight possible choices to write about. We control for this change in protocol by analyzing the data separately by year, as well as combined. Analyzing the combined data served as a check on temporal factors in the essays and tested popular notions of bigger data being better data [6]. See table 1 for the full breakdown of the data before preprocessing. The first step of preprocessing was to remove documents with fewer than 100 characters. In order to protect the identity of the university system which provided these data, we do not report the essay prompts provided to students.

|          | Students | Female | Male   | Essays  |
|----------|----------|--------|--------|---------|
| 2016     | 44,434   | 26,725 | 17,431 | 88,868  |
| 2017     | 48,702   | 29,710 | 18,727 | 194,808 |
| Combined | 93,136   | 56,435 | 36,158 | 283,676 |

**Table 1: Number of students and essays per year. Note that not every student reported gender.**

## 3.2 Classification Outcomes: Reported Household Income and Gender

We used applicant's reported household income (RHI) as a classification outcome. Clearly RHI is not an objective measure of a family's household income. We suspect some students reported RHI inaccurately. For example, some students reported RHI of 0 and others in the thousands of dollars rather than dollars. To address this, we filter out any student whose RHI is below $10,000. While some of these values may be accurate, we found that our ability to classify students goes up when we remove these students, indicating that the signal coming from these RHI values is unhelpful. Despite the limitations, RHI is an important variable because language variation along class and income lines has been well established in the sociolinguistic literature [3, 31]. Further, educational experiences and outcomes are shaped by a student's social class and tend to be strongly evident in linguistic and cultural practices [5]. We labeled essays as either above or below the median income for a given corpus. The median income was $42,000 for 2015; $44,000 for 2016; and $43,000 for the combined data. Students with RHI at exactly the median were labeled as below median income.

We used applicant's reported gender (RG) as a second classification outcome. Applicant RG was limited to "Male" or "Female" (1,349 essays were written by students that did not report gender and were excluded). While this binary is an incomplete measure of the gender spectrum, it still captures an important dimension of linguistic variation [15].

## 4 METHODS

We deployed several classification algorithms for prediction of RHI and RG. We sought to ascertain the extent to which admissions officers deploying computational reading techniques might face a tradeoff between the accuracy and interpretability of findings. Flexible blackbox models might offer the most accurate predictions of authors' demographic characteristics but provide little tractable insight for officers to inform their own readings of essays. Conversely, simpler techniques with lower degrees of predictive accuracy might prove more easily interpretable, and therefore warrant consideration as instruments for informing human reading.

After preliminary testing, we chose three models that represent increasing flexibility but decreasing transparency: Naive Bayes (NB), logistic regression (LR), and a deep neural model (DN). NB classifies based on word frequencies and is thus fairly interpretable, while DN is highly flexible but generally inscrutable apart from output statistics (eg. $f1$ score, precision). Each model has unique affordances that could provide important information to an admissions officer, whether they want to know the highest possible prediction accuracy or the words most associated with a label. We used zero-rule learning [9] to establish a baseline accuracy of 58% for gender and 50% for median income.

We used NLTK [32] to tokenize the documents and divided the data into 5 partitions for k-fold cross validation. Different selection procedures were used for each group of outcome variables which were described in section 3. Each model was trained separately on 6 different tasks: classify documents labeled by gender or median income on the 2015, 2016, and combined data. This was done for each of the five folds. The presented accuracies are the average $f1$ scores from the 5-fold cross-validation.

### 4.1 Multinomial Naive Bayes

Multinomial NB is a commonly used algorithm for text classification popular for its interpretability and accuracy. NB learns word frequency patterns between different document classes by calculating conditional probabilities [26]. Document-class probabilities are entirely generated by word frequencies for each class, making each word feature easily ranked and interpreted. The ease of interpretation would allow stakeholders that do not have computational backgrounds to understand the model's behavior. However, if the data is messy or complex, NB might miss patterns a human or more flexible algorithm could detect.

### 4.2 Logistic Regression

We used an LR classifier trained on unigram word counts per document with early stopping. We deployed other models that were more flexible than NB but not as easily interpretable (e.g. random forests, SVM, single layer perceptron); LR consistently outperformed them.

### 4.3 Deep Neural Model

Our DN architecture included one hidden layer of 150 nodes, and was trained using dropout [40] and $L_2$ regularization [11] to avoid overfitting. We used the sequential model from Keras, a linear stack of layers. We recognize that blackbox AI interpretability is an active area of research, but deep models still generally offer the least interpretability and highest flexibility.

## 5 RESULTS

### 5.1 Classification Accuracy

Table 2 reports the classification accuracies of the models. Each model was tested on the tasks of predicting RHI and RG on the 2015 data, the 2016 data, and the combined data (6 tasks total). Test accuracy was calculated using 5-fold cross-validation for each task.

We first note that, as expected, accuracy generally increases with model complexity, with DN outperforming LR, and LR outperforming NB. Despite being a relatively inflexible linear model, NB was able to achieve high accuracy on each task, indicating that word frequencies alone strongly encode these two demographic variables. We also notice that DN only slightly outperforms LR. Since the presence of nonlinearities barely improved accuracy, we believe that the problem itself is fairly linear. Though we tested a variety of deep models, it is possible that if our work focused solely on accuracy we could have achieved better results.

We also observe that all models are able to achieve higher classification accuracy on the 2015 data than on the 2016 data. This difference may be attributed to the fact that a single 2016 essay could be noisier than a single 2015 essay. Two factors from the essay format would cause this: first, the 2016 students chose four prompts from eight options while the 2015 students wrote two longer essays for the same prompts; second, applicants in 2015 were allowed 1,000 words for two essays and in 2016 had 350 words max per essay (average lengths: 480 and 312 words respectively).

|  | Reported Household Income | | | Reported Gender | | |
|---|---|---|---|---|---|---|
| Model | 2016 | 2017 | Combined | 2016 | 2017 | Combined |
| Naive Bayes | 65.02% | 62.85% | 63.23% | 73.27% | 70.04% | 70.91% |
| Logistic Regression | 67.87% | 65.27% | 66.46% | 79.37% | 75.14% | 77.16% |
| Deep Neural Model | 68.94% | 65.38% | 66.69% | 79.88% | 75.34% | 77.45% |

**Table 2: Test Classification accuracy for each model on each task.**

Although the models trained on the combined data used more documents than the models trained on the 2015 or 2016 data alone, they did not achieve the highest accuracies. This could point to temporal factors in the topics and narratives written by the applicants. This is understandable given that most of the 2016 essays were written during the 2016 presidential election. If the social and historical context produced essays with different levels of noise, this could also explain the lower accuracy seen in the models trained on the 2016 essays. And, as previously noted, the essay format changed from year to year and combining the data would have increased semantic and lexical variation.

Finally, it appears that gender is easier to predict than income, given that classification accuracies for RG are not only higher than the corresponding accuracies for RHI, but also represent a larger improvement over the zero-rule baseline. This suggests that word-usage is more indicative of RG than RHI, though an outcome less crude than median income could find more nuanced patterns. We explore differences between gender and income word frequency ratios in section 5.2 below.

## 5.2 Distinguishing Words: Naive Bayes Frequency Ratios

While NB achieved the lowest accuracy among our models on each task, it directly estimates the relative importance of features between classes and thus offers clear insights into its decision-making process. Specifically, for any word $w$, we can use the NB parameters to calculate a "frequency ratio"

$$FR(w) = \frac{P(w|\mathbf{0})}{P(w|\mathbf{1})}$$

which indicates how frequently word $w$ appears in writing with label **0** relative to writing with label **1**. Table 3 reports the 10 words most indicative for each label.

Table 3 allows us qualitative insight into the nature of student vocabularies in each of the four classes (above/below median RHI and Male/Female RG). While below-median RHI students write about immigration and financial burdens, above-median RHI students write about scouting and international travel. While male-identifying students write about video games and Rubik's Cubes, female-identifying students write about softball and ballet. Each half of Table 3 points to a specific difference in life experience between two groups. On the left side we observe a difference in *narrative* (immigration vs. travel), whereas on the right side, it is a difference in *activities* (male-dominated vs. female-dominated).

Finally, Table 3 yields insight into why, as mentioned in section 5.1, our algorithms were better at predicting RG than RHI. The FR values for RG go as high as **40**, whereas the highest values for RHI are only around **7**. This means that the degree to which a single word can signal gender is higher than the degree to which a single word can signal income. In our context, the two most discriminating words "softball" and "Latina" imply explicit gender segregation, exclusion, and identity. The same can not be said for the income features as they seemed to imply implicit segregation, exclusion, and identity.

## 6 DISCUSSION AND CONCLUSION

Holistic review is predicated on the notion that quantitative measures of academic accomplishment are not by themselves adequate bases for admissions decisions. Over time, admissions professionals have come to rely on qualitative sources of information, most notably written essays, to provide further material for evaluation. Yet the inclusion of qualitative sources in holistic review has long been controversial, as critics dispute the extent to which the assessment of qualitative materials accommodates human bias.

Advances in AI technologies have made it possible to "read" textual material statistically, blurring the distinction between quantitative and qualitative modes of evaluation. While much scholarship to date on the ethics of AI have focused on the negative risks associated with computational reading, relatively little attention has been paid to how such technology might be deployed to mitigate bias. Encouraged by recent judicial scrutiny of selective admissions, the research presented here sought to investigate the potential of computational reading as a check on implicit bias in holistic review.

Utilizing a corpus of a sort rarely available to researchers, we conducted computational readings of 283,676 essays submitted by 93,136 self-identified Latinx applicants to a selective public university system. We found that applicants' reported household income and gender could be predicted with high degrees of accuracy using widely available computational techniques. While data constraints obliged us to limit our inquiry to these two variables, our work provided proof-of-concept for utilizing computational reading to observe patterns on other demographic dimensions: ethno-racial identity, zip code, parental educational background, and high school type, for example. To the extent that such factors may be implicated in biased evaluation by human readers, they may be powerful tools for fulfilling Judge Burroughs' call for "a way to recognize when implicit bias might be affecting outcomes" in admissions decisions.

Computational readings such as those reported here might become routine practice in admissions offices as a means of sensitizing human readers to patterned variation in application materials and informing subsequent evaluation. Officers might then be able to more clearly see how patterned variation across an entire applicant pool can create hazards for biased readings – as when, as is common in employment hiring, applicants who report aptitudes and preferences more familiar to evaluators tend to enjoy preferential assessment [37]. We see promise for data auditing as a mechanism for sensitizing human evaluators to potential bias across a wide range of organizational contexts. Data auditing, either on its own or part of a datasheet [21], could alert humans about the potential for a dataset to smuggle in biases that AI would use for decision-making. Finally, data auditing could serve as a complementary framework to algorithmic auditing and bridge divides between researchers in AI and fields where fateful decisions are increasingly likely to be made or assisted by computational readings: education, healthcare, finance, and law, for example.

|  | Reported Household Income | | | | Reported Gender | | | |
|---|---|---|---|---|---|---|---|---|
|  | Below Median | | Above Median | | Male | | Female | |
| Rank | Word | FR | Word | FR | Word | FR | Word | FR |
| 1 | undocumented | 6.5 | eagle | 7.0 | baseball | 14.8 | softball | 41.5 |
| 2 | deported | 6.3 | scouting | 7.0 | skateboarding | 14.5 | latina | 39.3 |
| 3 | rent | 5.6 | switzerland | 6.9 | gaming | 12.4 | cheerleading | 12.2 |
| 4 | ets[2] | 5.4 | jewish | 6.8 | scouting | 12.0 | ballet | 11.9 |
| 5 | upward | 4.6 | paulo[3] | 6.7 | eagle | 11.9 | girl | 10.6 |
| 6 | bills | 3.7 | irish | 6.4 | chess | 9.7 | makeup | 9.7 |
| 7 | eld[4] | 3.6 | rowing | 6.0 | legos | 9.3 | girls | 8.6 |
| 8 | wage | 3.6 | scout | 6.0 | lego | 8.8 | female | 8.5 |
| 9 | gangs | 3.4 | brazilian | 5.9 | hardware | 8.1 | daughter | 7.9 |
| 10 | nephews | 3.3 | chile | 5.9 | rubik | 7.0 | cheerleader | 7.5 |

**Table 3: Words most indicative of each label according to the Naive Bayes models trained on combined data. Each word is listed along with its frequency ratio (FR), which refers to how many times more frequently students in one group use a word relative to students in the other group. For example, we can see that students with below-median RHI use the word "undocumented" 6.5 times as often as students with above-median RHI.**

Despite the promising potential of computational readings to inform and improve human-centered evaluation protocols, we must not be naive about the risks that inevitably come with the commingling of computation and human evaluation. First, we recognize that the same tools that can be used to inform evaluators to avert biased readings might also be used to identify groups for categorical discrimination or preference. If sensitive attributes like race and/or ethnicity are not handled carefully, computational approaches might do more harm than good. As with many sharp tools, computational reading techniques can cut in multiple ways.

Second, AI researchers and admissions officers conceive of fairness and bias in different and important ways. AI researchers tend to be concerned with fairness and bias at the population level, and worry when patterned evaluative outcomes do not approximate population demographics [45]. By contrast, admissions officers tend to emphasize fairness of evaluation for individual applicants [2]. We should not expect that these two ethical emphases can be easily or seamlessly integrated. Critical, ongoing, collegial engagement between admissions and AI professionals would be necessary to optimally connect these different ethical imaginaries.

Third, some categories of applicants that might matter for admissions officers – "distinctive individuals" or "future leaders," for example – might not be amenable to easy identification for computation. Further, even if methods were developed to identify or group students fitting any number of criteria, it is not mathematically possible to meet all definitions of fairness; doing so might even harm groups of people we hope to protect by enforcing fairness constraints [10]. The process of combining algorithms and humans in decision making processes also requires more understanding before data auditing in college admissions is feasible [23].

Finally, it is important to note that classifications were not perfect, with the best models achieving $f1$ scores of 69% (RHI) and 80% (RG). A less than perfect prediction raises legitimate questions regarding the utility of our approach for data auditing in admissions. Incorrect classification may itself lend bias to subsequent human reading. Our work might also be misconstrued as providing insight on the "quality" of the essays or an invitation for such analyses. Grace, voice, wit, and insight remain elusive to computational reading. On those things our work is silent.

Nevertheless, there is great value in seeing patterns in data before they are transformed into biased judgments downstream. In the same vein, our encouragements regarding computational readings in selective admissions are analogous to ones made about the importance of understanding genetic data for improving health outcomes in medical care [28]. If decision makers know in advance the patterned variation in their data that could leak into diagnoses or evaluations, they can be better equipped to proactively address biases.

## 7 ACKNOWLEDGMENTS

Thank you to our data providers and our colleagues at the Stanford Graduate School of Education for their helpful feedback.

---

[2] The Educational Talent Search (ETS) is a federal program designed to help low-income students enroll in college.
[3] English Language Development
[4] Presumably from São *Paulo*, the largest city in Brazil